\documentclass[12pt,preprint]{aastex}

\shorttitle{GS2000+25}
\shortauthors{K. Terada, S. Kitamoto \& H. Negoro}

\begin{document}

\title{
Energy Spectra and Normalized Power Spectral Densities of X-Ray Nova
 GS 2000+25}

\author{Kentaro Terada}
\affil{Department of Earth and Planetary System Science, Hiroshima University,
1-3-1 Kagamiyama, Higashi-Hiroshima, Hiroshima 739-8526}
\affil{terada@sci.hiroshima-u.ac.jp}

\author{Shunji Kitamoto}
\affil{Department of Physics, Rikkyo University, 
3-34-1 Nishi-Ikebukro, Toshima-ku 171-8520}

\author{Hitoshi Negoro} 
\affil{The Institute of Physical and Chemical Research,
2-1 Hirosawa, Wako, Saitama 351-0198}

\and

\author{Sayuri Iga}
\affil{Hitachi Ltd. Totsuka-ku, Yokohama 224-0817}
\affil{ (Received 1994 April 4; accepted 2002 June 28) }

\begin{abstract}

The X-ray energy spectra and Normalized Power Spectral Densities (NPSDs)
of an X-ray nova, GS 2000+25, were investigated. The X-ray energy
spectra of the source consist of two components: a hard component,
which can be represented by a power-law, and an ultra-soft component,
represented by radiation from an optically-thick
accretion disk (the disk component). In a model in which the power-law
component is the Compton-scattered radiation, it is found that the
temperature of the incident blackbody radiation to the Compton cloud
decrease from 0.8 keV to 0.2 keV according to the decay of the intensity, which coincides with  that of the
inner accretion disk. When the source changed from the high-state to
the low-state, both the photon index of the power-law component (or
Compton $y$-parameter) and the NPSD of the hard component dramatically
changed as did GS 1124-683. That is, the photon index changed from
2.2--2.6 to 1.7--1.8 and the absolute values of the NPSDs at 0.3 Hz of
the hard component in the low-state became about 10-times larger than those of
the hard component in the high-state. These
X-ray properties were similar to those of other black-hole candidates,
such as Cyg X-1, GX 339-4, and LMC X-3.  
\end{abstract}

$Key$ $words:$ accretion disk --- black holes --- stars: individual: (GS
2000+25)--- stars: novae --- X-rays: binaries

\section{Introduction}

X-ray novae are classified by their early observations of the energy
spectra into ``hard'' ($kT >$ 15 keV) and "soft" ($kT <$ 15 keV) classes
(Cominsky et al. 1978). White et al. (1984) proposed the third class
of the transients that showed ``ultra-sof'' ($kT <$ 3 keV) energy
spectrum. The persistent X-ray sources which show such ultra-soft
spectra include a black-hole candidate X-ray binary (BHC-XB), such as
Cyg X-1 in the high state and LMC X-3 (White, Marshall 1984). While
not all ultra-soft sources may contain accreting black-holes, this
empirical classification has proved its predictive power in the case
of the ultra-soft transient A0620-00, which McClintock and Remillard
(1986) subsequently identified as a BHC-XB based on radial
velocity measurements. GS 2000+25 and GS 1124-683, which are
mentioned in this work, belong to this class.

The most reliable criterion for a black hole composing an X-ray binary
(XB) is that the lower-mass limit must exceed 3 $M_\odot$. The current
theories predict that a neutron star with more than 3 $M_\odot$ cannot remain
stable, and will collapse into a black hole (see, e.g., Baym, 
Pethick 1979). It is known that there are at least ten X-ray binary
systems, in which the compact object has a lower mass limit of more
than 3 $M_\odot$ (e.g. Tanaka, Shibazaki 1996; Barret et al. 1996).

Mainly from observations of Cyg X-1, the following features of X-ray
emissions have been considered to be possible black-hole signatures:
(1) short-term variations of the X-ray emissions (so-called
flickering), (2) an ultra-soft energy spectrum in the high X-ray
intensity state (the high state), (3) a hard (power-law type) energy
spectrum in the low X-ray intensity state (the low state), and (4)
bimodal behavior between these two states. Tanaka
(1989) reported that almost all of the observed energy spectra of
BHC-XBs could be represented by a two-component model consisting
of a multi-colored disk blackbody (Mitsuda et al. 1984) and 
power-law shape radiation. Subsequently, Miyamoto et al. (1991),
Takizawa (1991), Ebisawa (1991), and Ebisawa et al. (1994) reported
that the observed energy spectra of BHC-XBs were able to be represented by
a two-component model consisting of a multi-colored disk
blackbody and power-law radiation with a smeared edge
absorption. (For the reviews of the BHC-XBs, see Oda 1977; Liang, Nolan 1984; Tanaka 1989; Gilfanov et al. 1999; Kitamoto 1999, and the references
therein.)

The structure of the accretion disk depends on the accretion rate
(e.g. Frank et al. 1992; Kato et al.
1998). The standard disk (Shakra, Sunyaev 1973) is one of the
stable solutions in the case of an optically-thick accretion
disk. Another stable solution in an optically-thick disk is known as
advection-dominated accretion flows (ADAFs), or a ``slim disk''
(Abramowicz et al. 1988, 1995; and references therein), where the X-ray
spectrum becomes harder than that of the standard accretion disk (e.g.
Watarai et al. 2001).  In the case of a low accretion
rate, the accretion disk becomes optically thin.  There are two
solutions: one is the gas pressure being supported and radiately cooled,
which is thermally unstable. Another is gas pressure supported by being
advectively cooled; this is stable. An original idea concerning the
transition from an optically-thick disk to an optically-thin disk
for the bimodal behavior between the high and low states of the
BHC-XBs was proposed by Ichimaru (1977). Since X-ray novae must
change the accretion rate with many orders of magnitude, it is the best
target to investigate the behavior of the accretion rate as a function
of the mass-accretion rate.

Rapid time variations of X-rays have also been well investigated using
the power spectral density (PSD; Terrell. 1972; Weisskopf et al. 1975; Nolan et al. 1981) and other methods (see for instance, Oda
1977; Liang, Nolan 1984; Makishima 1988; and the references
therein). In addition, the Normalized Power Spectral Densities (NPSDs)
have been used to compare the PSD of Cyg X-1 at different occasions
(Belloni, Hasinger 1990), and among different sources (Miyamoto et al. 
1991, 1992, 1993, 1994; Kitamoto et al. 1992a, b), which are the
power spectral densities normalized by its intensity, where the
Poisson noise due to statistical fluctuations is subtracted. Phase
lags have also been measured between the time variations of different
energy X-rays (van der Klis et al. 1987; Miyamoto et al. 1988, 1991,
1992, 1993, 1994).

Miyamoto et al. (1992) reported that, in the low state of BHC-XBs the
shapes and the absolute values of the NPSDs and phase lags of
time variations between different energy X-rays were
similar. Moreover, they reported that the shapes and the absolute
values of the NPSDs in the high state of BHC-XBs were quite different
from those in the low state; i.e. the absolute values of the NPSDs  in the high state were smaller than those in the low state by
factor of about 10--1000 at the frequency range above 0.1 Hz (Miyamoto
et al. 1993).

Miyamoto et al. (1994) pointed out, for GS 1124-683 in
the high state, that in the 1.2--37 keV energy range, the two energy
spectral components (i.e. a power-law component and a disk blackbody
component) have power spectral densities (PSD) of their own
characteristic shapes; the PSD of the power-law component has a
flat-top shape (FT noise function) at frequencies of less than about 2
Hz, and the PSD of the disk blackbody component has a power-law
shape (PL noise function). In this state, the FT noise functions
normalized by the power-law component are almost the same as in the cases
when the photon counts fraction of the power-law component (PLF) is
larger than about 20\% of the total photon counts in 1.2--37 keV, and
the PL noise functions normalized by the disk blackbody component are
also almost the same when the PLF is less than about 10\%. Moreover, it
is found that the so-called high state can be divided into two kinds of
sub states: a very high state and a high quiet state. In the high
quiet state, the ultrasoft component dominates the X-ray spectrum, and
the time variation is ``quiet''. On the other hand, in the very high state,
the ultrasoft component dominates, but the power-law component also exists substantially, and the total time variation is relatively high. These completely
correspond to the ``high state'' and the ``very high state'' by van der Klis
(1994a, b). These studies on the rapid time variations and components of X-ray emission must be relate to the structure of the accretion disk. The phenomenological behavior of the BHC-CBs has been reviewed by Belloni (2001).

A bright transient X-ray source in the constellation Vulpecula, GS
2000+25, was discovered with the All Sky Monitor (ASM, Tsunemi et
al. 1989a) onboard the Ginga satellite (Makino. 1987)
on 1988 April 23 (Tsunemi et al. 1989b). After the discovery of GS
2000+25, the source was observed from 1988 April 30 to 1988 December 16
with the Large Area Counter (LAC; Turner et al. 1989) onboard
the same satellite. The observed maximum X-ray intensity was
approximately 30 count s$^{-1}$ cm$^{-2}$ (1--20 keV), and the X-ray energy spectra
observed at the time of the flare were of the ultra-soft type, and the
light curve observed during the decay phase was similar to that of the
BHC-XB, A0620-00.

After some trials to determine the mass of the compact star in
GS2000+25 (Charles et al. 1991; Callanan et al. 1991; Mineshige et
al. 1992a), the orbital motion of the binary system of GS2000+25 was
precisely determined based on an optical observation with the
W. M. Keck 10 m telescope (Filippenko et al. 1995). The obtained mass
function of 4.97$\pm$0.10 $M_\odot$ indicates that the plausible mass of the
compact star is 5.9--7.5 $M_\odot$ assuming that the mass of the
companion star is 0.4--0.7 $M_\odot$, and that the inclination of the
orbital plane is 67.5--80 degrees. Casares et al. (1995) also determined
a mass function of 5.02$\pm$0.46 $M_\odot$ and implied the probable mass
of the compact star is (7.0--7.7)$\pm$0.5 $M_\odot$ based on the
spectroscopic detection of the secondary star. Beekman et al. (1996)
reported that mass of the compact object lies between 4.8 and 14.4
$M_\odot$ from the infrared light curve. Callanan et al. (1996) reported
that the mass of the compact star is 8.5$\pm$1.5 $M_\odot$ assuming that the
companion star is a K dwarf. All of these results suggest that the mass
of the compact star is more than 3 $M_\odot$, indicating that GS 2000+25 is
a BHC-XB.  Garcia et al. (2001) reported the significant detection of
GS2000+25 by a Chandra observation, which shows a luminosity of
2.4 $\times$ 10$^{30}$ erg s$^{-1}$. This low luminosity also
suggests that the compact star is a black hole.

The X-ray emission from GS 2000+25 decreased over a period of eight months, changing
in the absolute values and appearance of the energy spectra and the
NPSDs, the features of which remind us of the so-called bimodal behavior
of the BHC-XBs, such as Cyg X-1 and GX 339-4 and GS 1124-683. In this
work, the time evolution of the energy spectra and the NPSDs were
investigated in detail; the correlation between them is discussed. These results are discussed according to the accretion-disk structure.  In
this study, the classifications of the state (low state, very high
state and high quiet state) are pursuant to those by Miyamoto et al.
(1993, 1994).

\section{Observation}

After the discovery of GS 2000+25 with the ASM on 1988 April 23, GS
2000+25 was subsequently observed with the Large Area Counter (LAC;
Tunner et al. 1989) intermittently from 1988 April 30 to 1988 December
16. The start and end times, observation modes, energy channels and
time resolution of the observation are listed in table 1. Days from 0h
0m 0s (UT) on 1988 April 28, when the X-ray flux observed with the ASM
was maximum, are also shown in parenthesis in the table. Hereafter,
the origin of time is defined as 0h 0m 0s (UT) on 1988 April 28 and
the days from then will be called ``days after the peak'', although we do not have any data between April 23 and April 28. In order to investigate the energy spectra, MPC-1 and MPC-2 data,
of which the number of energy channels  was 48, were used. MPC-3 data, of which
the time resolution was 7.8 ms, were used in order to investigate
rapid time variability of X-rays.

A further observation, not described in this work, was carried out with the
LAC on 1989 May 15,
about 12 months after the peak of the outburst, but the source was
below the LAC detection limit of $\sim$ 0.3 mCrab (Mineshige et al. 1992b).

\section{Analysis and Results}

\subsection{$Light$ $Curve$}

Figure 1 shows light curves (observed photon counts) of GS 2000+25 in
the 1--6 keV and 6--20 keV energy ranges and its hardness ratio,
observed with the ASM. After the initial detection on 1988 April 23,
the X-ray intensity of this source reached about 12 Crab in the 1--6 keV
energy range on April 28, and decreased exponentially thereafter (Tsunemi
et al. 1989b). It flared up slightly again approximately 70 d and
132 d after the peak in the 1--6 keV energy range. The decay time
constant was approximately 30 d until 61.7 d after the peak. In the
6--20 keV energy range, GS 2000+25 flared up again 132 d after the
peak, and the decay time constant was about 20 d until 61.7 d after the
peak. That is, hard X-rays decreased faster than soft X-rays in the
early phase of the outburst (until 127 d after the peak), as shown by
the hardness ratio. The hard X-rays (6--20 keV) were not the main
component at the beginning of this outburst, as long as GS 2000+25 was
observed with the ASM. The hardness ratio suddenly increased on the 129 d
after the peak.  As shown in figure 1, slight flarings 70 d
and 132 d after the peak are different from the behavior of their X-ray energy spectra.

The light curve obtained with the LAC is shown in figure 2. In the
1.7--37 keV energy range, the decay time constant was 33.9 d until
1988 December 16; 232 d after the peak. Due to intermittent data
obtained with the LAC, the re-flares, which were observed in the ASM,
were not clear.

\subsection{$Energy$ $Spectra$ $of$ $X-rays$}

The energy spectra of X-rays observed with the LAC (1.7--37 keV) over eight
months are shown in figure 3. 
For these data, an aspect correction and background subtraction were performed, but a correction of the detector response was not applied.
It should be noted concerning figure 3 that soft
X-rays below 8 keV decreased with time, but hard X-rays above 8
keV increased once on September 8 (133 d after the peak).

An attempt was made to fit these X-ray energy spectra with three kinds of
two-component models; an ultra-soft component and a hard component. To
simulate the ultra-soft component, the multi-colored disk blackbody
model (Mitsuda et al. 1984; hereafter, termed``the disk component'')
was used in all 3 models. For the hard component, the following 3
models were used: (model-1) a power-law component, (model-2) a
power-law component with a smeared-edge (Takizawa 1991; Ebisawa 1991;
Ebisawa et al. 1994), and (model-3) a Comptonized blackbody component
(Nishimura et al. 1986). The best-fit parameters of the energy spectra
based on these three models are shown in tables 2, 3 and 4. The error bars
quoted here are the 90\% CL. For fitting the
parameters of the disk component, the innermost disk radius ($R_{\rm in}$)
multiplied by $\sqrt{\cos i}$, assuming the distance to the source to
be 1 kpc, and the innermost temperature of the disk ($kT_{\rm in}$) are
listed in the three tables, where $i$ is the inclination angle of the
disk. In tables 2 and 3, the photon index of the power-law component is
shown. In table 4, instead of the photon index, parameters of the
Compton scattering process are shown, such as $kT_{\rm bb}$,
$kT_{\rm electron}$, and $\tau$, which are the temperature of incident
blackbody radiation into the Compton cloud, the temperature of the
Compton cloud and the radius of the Compton cloud in Thomson
scattering depth. The normalization factor of the Comptonized
blackbody in table 4 is the surface area of the blackbody radiation
incident to the Compton cloud. The normalization factors in tables 2
and 3 were calculated assuming the distance to the source to be 1 kpc. In
table 4, the Compton $y$-parameter, which is defined as (4
$kT_{\rm electron}$/$m_{\rm e}$ c$^2$)$\times$ Max($\tau$,$\tau ^2$), is shown
while assuming non-relativistic thermal distributions of the electrons, in order
to evaluate the property of the hard component. The fluxes of the disk
and the hard component in the tables are the energy fluxes of X-rays
from 1.7 keV to 37 keV. The observed photon count fraction of the hard
component to the total counts (the count fraction of the power-law
component; PLF) and the reduced $\chi ^2$ values of the fitting are
also given in these tables. Here, the PLF is defined as
$I_{\rm hard}$/($I_{\rm hard}$+$I_{\rm disk}$), determined by each energy spectral model convolved
by the detector response, where $I_{\rm hard}$ and $I_{\rm disk}$ are the observed X-ray
counts of the hard and disk components in the 1.7--37 keV energy
range, respectively.

Among the above models, the model 2 gives the smallest $\chi ^2$ values. The
energy spectra with the best-fit curves of model 2 and its
residual are shown in figure 4, and the time evolution of the fluxes
of the two X-ray components derived by model 2 is shown in figure
5 together with those of GS 1124-683 (Miyamoto et al. 1993). The evolution
of the various parameters of models 2 and 3 is shown in figures 6 and
7, respectively. Comparisons of these results with those from
GS 1124-683 and further discussions about these two sources are
presented in subsection 4.1.

In all models, the innermost temperature ($kT_{\rm in}$) of the disk near the peak X-ray
flux was 1.1--1.2 keV, and then smoothly decreased down to about 0.3 keV
as the X-ray flux decreased. At the same
time, the innermost disk radius ($R_{in}$$\sqrt{\cos i}$) remained
almost constant ($\sim$10$^6$ cm).

From April 30 (2 d after the peak) to May 7 (9 d after the peak),
the photon counts in the 20--37 keV high-energy range are small and
statistically poor, as shown in figure 4. Therefore, it is difficult
to derive any conclusions relating to the hard component at this
time. From June 1 (34 d after the peak) to December 9 (225 d after the
peak), the photon indices of the power-law component were larger than
about 2.2, and suddenly decreased to about 1.8 on December 16 (232 d
after the peak) in the case of models 1 and 2, as shown in tables 2, 3
and figure 6. For model 3, the Compton $y$-parameter was smaller
than 0.5 until December 9, and then suddenly increased to about 1.0 on
December 16, as shown in figure 7.

The temperature of the incident blackbody radiation to the Compton
cloud gradually decreased from 0.8 keV on June 1 (34 d) to 0.2 keV on
December 16 (232 d). The cooling of the incident blackbody temperature
was similar to that of the inner disk ($kT_{\rm in}$). In all models, the flux
of the hard components reached the maximum on June 1 (34 d after the
peak).

On December 16, the PLF was about 90\% and the X-ray energy spectrum
was  a power-law type with a photon index of about 1.7--1.8. This is
similar to the energy spectra of Cyg X-1 and GS 1124-683 in the
low state (Miyamoto et al. 1992, 1994). Moreover, as mentioned
in subsection 3.3, the NPSD on December 16 is similar to those of Cyg X-1 and
GS 1124-683 in the low state (Miyamoto et al. 1992, 1994). Therefore,
it can be regarded that GS 2000+25 on 1988 December 16 was in the
low state. 
Recently, the intermediate state has been reported for CygX-1 in 1996, GX339-4 in 1998
(Belloni et al 1996, 1999; Esin et al. 1998). According to a similarity of the photon indices and the short-term variability of X-rays, as Belloni et al. (1996) have already suggested, they resemble the
very high state of GS1124-68 in 1991 and GX339-4 in 1988 described in the Miyamoto et al. 
(1993, 1994). Miyamoto et al. (1993) originally called this state of GS1124-68 a high-to-low 
transition state. Between 130 d and 225 d in the case of GS2000+25, we could apparently
recognize a hardening of the spectrum, but still a substantial soft component was
observed. Therefore, this term can be considered to be a kind of
transition phase, which is different from the case of GS1124-683.

\subsection{$Normalized$ $Power$ $Spectral$ $Densities$ $(NPSDs)$}

In order to investigate any short term variability of the X-rays from GS
2000+25, the Normalized Power Spectral Density function (NPSD,
Miyamoto et al. 1991, 1992, 1994) was calculated. This function is
suitable to compare the time variations of different sources, and also the
same source at different intensities. Miyamoto et al. (1992) pointed
out that the shapes and values of NPSDs of various BHC-XBs in the
low state were very similar to each other, except at frequencies below
about 0.2 Hz, which suggested that various BHC-XBs in the low state
had the same mechanism of X-ray time variability. Miyamoto et al. (1993) also
pointed out that the shape and values of the NPSDs of GX 339-4 and GS
1124-683 in the high state were similar.

The NPSD is the power spectral density normalized to its intensity,
where the Poisson noise due to statistical fluctuations is subtracted,
which is defined as the square of the ratio of the root-mean-square
amplitude of the two-sided power spectral density in unit frequency
band (Hz$^{-1}$) to the total signal photon number as shown below (Miyamoto
et al. 1994):

\begin{eqnarray}
{\rm NPSD}_k = \frac{(a_k^2 +B_k^2- \frac{1}{n} \sigma ^2)\times T}{(\overline{x}-B)^2}, \\
a_k = \frac{1}{n}\sum^{n-1}_{j=0} x_j \cos (\frac{2 \pi k j}{n}) \nonumber , \\
b_k = \frac{1}{n}\sum^{n-1}_{j=0} x_j \sin (\frac{2 \pi k j}{n}) \nonumber , \\
k = 0, 1, 2,..... ,\frac{n}{2} \nonumber, 
\end{eqnarray}

\noindent where $x_j$ is the number of counts in the $j$th bin of $n$
consecutive time bins, $\overline{x}$ (counts per bin) is their mean
value and their background $B$ (counts per bin), $\sigma ^2$ is the
estimated power due to the Poisson statistics of the data
(=$\overline{x}$), $T$ is $n \Delta T$ (s) ($\Delta T$ is the bin
length), and the frequency $f$ (Hz) is given by $f=k/T$ (Miyamoto et
al. 1991, 1992, 1993, 1994; Kitamoto et al. 1992a, b).

The NPSDs of GS 2000+25 at three different PLF are shown in figure
8. The numerals in each diagram are the PLF, the observed date and the days after
the peak. Typical NPSDs of GS 1124-683 and of Cyg X-1 are also shown in
this figure; the thin solid line is the NPSD of GS 1124-683 obtained
on 1991 January 11 (PLF=0.80). The dotted line is the NPSD of GS
1124-683 obtained on February 26 (PLF=0.066). The broken line is the
NPSD obtained on June 13 and July 22 when GS 1124-683 was in the
low state (PLF=0.95 and 0.99). The thick solid line is the NPSD of Cyg
X-1 in the low state (1990 May 10).

From figure 8, it was found that the NPSD on December 16 coincided
with the NPSDs of GS 1124-683 (broken lines) in the low state, which
together with the energy spectrum mentioned in subsection 3.2,
clearly indicates that GS 2000+25 is in the low state on December
16. It was also found that both the energy spectra and the NPSD of GS
2000+25 and GS 1124-683 also had similar shapes in the high state.

Miyamoto et al. (1994) showed that the two energy spectral components
had their characteristic PSD noise functions in the high state of GS
1124-683; the flat-top shape noise for the power-law component and the
power-law shape noise for the disk component, respectively. Applying
the same noise functions to GS 2000+25, we obtained the following results:

Following Miyamoto et al. (1994), the NPSDs of GS 2000+25 were fitted
with a model that consisted of two characteristic power spectrum
densities (PSDs), PSD$_{\rm power law}$ and PSD$_{\rm flat top}$, which were represented by
a Lorenzian function of zero central frequency. That is,

\begin{eqnarray}
 {\rm NPSD}_{\rm observed}(f) &=& {\rm PSD}_{\rm power law}(f)+{\rm PSD}_{\rm flat top}(f)\nonumber \\
	&=& A_1 f^{-k} + A_2 \frac{(\Gamma/2)^2}{f^2 + (\Gamma/2)^2} , 
\end{eqnarray}

\noindent
where $f$ is the frequency, $\Gamma$ is the FWHM of the Lorenzian, $A_1$ and
$A_2$ are constants.

In some cases, these two-component models are not necessarily required from a statistical point of view, because the F-test shows no significance for the two components.
However, judging from an analogy with GS1124-683, we assumed
that the NPSDs could be expressed by a combination of two
characteristic shape functions, the power-law type and the flat-top
type; furthermore, a QPO was necessary in the case of September 8.

At first, the values of $k$ of the power-law component, $A_1 f^{ -k}$,
were investigated by applying a single power-law model  to the NPSDs on May 2, May 4, May 7, and May 8, when the PLF
was smaller than 0.03. The values of $k$ and its reduced $\chi ^2$ values are
given in table 5. The $k$ value of 0.807 on May 7 will be used as a
fixed value, because the PLF is the smallest and the NPSD in May is
considered to be the time variabilities of the disk component.

Next, in order to determine the value of $\Gamma$, fitting the
NPSDs on September 8, October 18 and November 5 with $A_1 f^{-0.807}+A_2 \frac{(\Gamma/2)^2}{f^2+(\Gamma/2)^2}$ was tried. In the case of September
8, another Lorenzian function with a center frequency, $f_{\rm QPO}$ of about
2.27 Hz was added in order to simulate the Quasi Periodic Oscillation
(QPO). The values of $\Gamma$ and a reduced $\chi ^2$ are given in Table 6. The
reduced $\chi ^2$ value on September 8 is not small enough, because the NPSDs
at a low-frequency range less than 0.06 Hz tend to deviate from the sum
of the two functions. At a higher frequency range, especially above
about 0.3 Hz, the sum of the two functions can represent the observed
NPSDs well. Thus, the $\Gamma$ value of 8.96 Hz on September 8 will be used
as a fixed value, because the PLF obtained on September 8 was the
largest, except for December 16. These values were similar to those of
GS 1124-683, where $k$ was 0.7 and $\Gamma$ was about 8 Hz (Miyamoto et al. 
1994).

At last, fitting the NPSDs with equation(2) was tried, while fixing $k$ and
$\Gamma$ values to be 0.807 and 8.96, respectively.  
In figure 9, the NPSDs of GS 2000+25 over eight months and their fitting
results are shown. The numerals in each diagram are the observed date,
the days after the peak, and the PLF. The dotted line, the broken line
and the thin line are PSD$_{\rm power law}$, PSD$_{\rm flat top}$, and their sum. The thick line in the figure for December 16 is the NPSD of Cyg
X-1 in the low state on 1990 May 9. The fitting results of the NPSDs and
the time resolution of each datum are given in table 7.

\subsection{$Relation$ $between$ $the$ $NPSDs$ $and$ $the$ $Power-law$ $Fraction$ $(PLF)$}

The relation between the values of the NPSDs at 0.3 Hz and the PLFs for GS
2000+25 was investigated. Here, the PLF values derived by model 2 and
the NPSD values at 0.3 Hz by equation (2) with two fixed values of
$k$ =0.807 and $\Gamma$ =8.96 were used. The relation is shown in
figure 10, together with the results of GS1124-683 (Miyamoto et
al. 1994). Here, we note that the value of 0.3 Hz is not a single data
point, but a representative value which was derived from the best-fit
model.

As shown in figure 10, a notable correlation between the values of
NPSDs and the PLF is recognized. That is, the larger is PLF, the
larger is the value of NPSD. More noteworthy is that this correlation
between the PLF and the value of NPSD of GS 2000+25 (filled circle) is
very similar to that for GS 1124-683 (open square, Miyamoto et
al. 1994). This reveals that the two sources
show not only the same energy spectral type, but also the same time
variability on the way from the outburst to the low state, even though
details of the evolution of the two sources from the outburst to the
low state are different, as shown in figure 5, and will be mentioned in
subsection 4.1.

If it is assumed that each of the hard (power-law) and disk
components has their characteristic values of the NPSD at 0.3 Hz, and
that these are independent from each other, the observed NPSD (0.3)
should be expressed as a function of the PLF by (Miyamoto et al. 1994):

\begin{eqnarray}
{\rm NPSD}(0.3) &=& \frac{({\rm NPSD}_{\rm hard}(0.3)\times I_{\rm hard}^2 + {\rm NPSD}_{\rm disk}(0.3)\times I_{\rm disk}^2)}{(I_{\rm hard}+I_{\rm disk})^2} \nonumber \\
      &=& {\rm NPSD}_{\rm hard}(0.3)\times {\rm PLF}^2 + {\rm NPSD}_{\rm disk}(0.3) \times (1-{\rm PLF})^2 , 
\end{eqnarray}

\noindent where NPSD$_{\rm hard}$(0.3) and NPSD$_{\rm disk}$(0.3) are the
NPSDs of the hard and disk components at 0.3 Hz normalized by the
hard and disk components, respectively. In figure 10, the most
probable fitting results of this formula to the NPSD values at 0.3 Hz
derived from equation (2) with two fixed values, except for the data in
the low state, are shown by the solid line for GS 2000+25 and the
broken line for GS 1124-683 (Miyamoto et al. 1994). The values of
NPSD$_{\rm hard}$(0.3) and NPSD$_{\rm disk}$(0.3) of GS 2000+25 are (1.59
$\pm$ 0.06) $\times$ 10$^{-3}$ and (2.4$\pm$1.0) $\times$ 10$^{-5}$,
respectively (reduced $\chi ^2$=3.7). Also, those of GS 1124-683 are (9.4$\pm$0.9)$\times$
10$^{-4}$ and (1.36$\pm$0.53)$\times$ 10$^{-5}$, respectively
(Miyamoto et al. 1994). These curved lines approximately show the
observed values of the NPSDs, although there are several
exceptions. This suggests that the NPSD$_{\rm hard}$($f$) and the
NPSD$_{\rm disk}$($f$) have almost similar characteristic values of about
10$^{-3}$ and 10$^{-5}$ at 0.3 Hz, respectively, on various occasions and
various BHC-XBs in the high state. Moreover, the values of
NPSD$_{\rm hard}$(0.3) in the low state are larger than those of
NPSD$_{\rm hard}$(0.3) in the high state by a factor of about 10 at 0.3
Hz. This is common to both GS 2000+25 and GS 1124-683. It is of
interest to note that in the high state both of the values of the
NPSD$_{\rm hard}$(0.3) and the NPSD$_{\rm disk}$(0.3) of GS 2000+25 are larger
than those of GS 1124-683 by a factor of 1.7--1.8. Both the hard and disk components of GS 2000+25 should have a higher variability by a
factor of 1.7--1.8 than those of GS 1124-683. One can not shift the
solid line to the lower side by adding a constant DC X-ray flux
component to GS2000+25.

\subsection{NPSD$_{\rm power law}$($f$) $and$ NPSD$_{\rm flat top}$($f$)}

It is described in subsection 3.3 that the observed NPSDs can be represented by
two components of the shape functions of PSD$_{\rm power law}$($f$) and
PSD$_{\rm flat top}$($f$). In subsection 3.4, it is shown that the NPSD consists of two
components, the NPSD$_{\rm hard}$ and the NPSD$_{\rm disk}$, which have almost
constant characteristic values at different PLF, and even in different
sources. It is natural to assume that PSD$_{\rm power law}$($f$) and
PSD$_{\rm flat top}$($f$) correspond to the PSD of the disk and the hard
component, because when the PLF is negligibly small (May 7) the NPSD
shows the power-law shape, and when the PLF is largest in the
high state the NPSD shows the flat-top shape (September 8). That is,
the PSD$_{\rm power law}$($f$) and the PSD$_{\rm flat top}$($f$) normalized by the
intensities of the disk component and the hard component,
respectively, should correspond to the NPSD$_{\rm disk}$($f$) and the
NPSD$_{\rm hard}$($f$). With this assumption, the values of the PSD$_{\rm power law}$($f$)
and the PSD$_{\rm flat top}$($f$) at 0.3 Hz, normalized by the intensities of the
disk component and the power-law component with a smeared edge
(model-2), respectively, are shown in figure 11. The data of GS
1124-683 are also shown in this figure (open square; Miyamoto et
al. 1994).

In the case of GS 2000+25 in the high state (PLF $<$ 0.61), the mean
value of the PSD$_{\rm power law}$ at 0.3 Hz normalized by the disk
component is (2.7$\pm$0.6)$\times$10$^{-5}$, assuming that the values of
the NPSD$_{\rm disk}$ are constant (the upper panel in figure 11; reduced
$\chi ^2$=1.7). On the other hand, the mean values of the PSD$_{\rm flat
top}$ at 0.3 Hz, normalized by the power-law component with a smeared
edge (model-2), is (9.7$\pm$1.8)$\times$10$^{-4}$ in the high state
(PLF $<$ 0.61), assuming the values of the NPSD$_{\rm hard}$ are constant
(the lower panel in figure 11; reduced $\chi ^2$=1.2). These are
almost consistent with values of the NPSD$_{\rm disk}$(0.3) and
NPSD$_{\rm hard}$(0.3) in subsection 3.4. It is also shown that the values of the
PSDs at 0.3 Hz, normalized by each component, are almost constant. Thus,
it is confirmed that two energy spectral components have characteristic
NPSDs in the high state, and that the correlation between the PLF and the
value of the NPSDs at 0.3 Hz (in subsection 3.4) are explained by the
characteristic NPSDs normalized by the two energy spectral components
and their independent variabilities from each other (in subsection 3.5).

It was reported that the normalized values of the PSDs of each
component were not constant in the case of GS 1124-683, as shown in
figure 11 (Miyamoto et al. 1994). The property of GS 2000+25 is
different from that of GS 1124-683, as follows. When the PLF is above
0.1, the normalized PSD$_{\rm power law}$ of GS 1124-683 tends to increase, but
this tendency is not seen in the case of GS 2000+25. This may be due to the poor statistics of the data of GS2000+25.  Especially, the
value of the NPSDs of GS 1124-683 at a PLF of 0.80 is larger than that of GS
2000+25 at a PLF of 0.61 by two orders of magnitude. This difference may
result from the difference in the evolution stage of the hard
component, because in GS 1124-683, the datum of a PLF of 0.8 is in the
initial phase of the nova (5 d before the X-ray maximum). On the other
hand, the datum of the PLF=0.61 of GS 2000+25 is after a re-increase
of the hard component (in the high to low transition phase; Miyamoto
et al. 1993).
	
\subsection{$Phase$ $Lags$}

Another way to investigate the short-term variability of the X-ray
intensities is the phase lags between the time variations of different
energy X-rays (van der Klis et al. 1987; Miyamoto et al. 
1988). Miyamoto et al. (1992, 1993) reported that the phase lags
between the short-term  variations of different energy X-rays from
BHC-XBs in the low state were similar. They also showed that just
after the flaring up of GS 1124-683 and GX 339-4, a peculiar large
peaked phase lag was observed. However, in the case of GS 2000+25,
the results are inconclusive owing to the large error in the data.

\section{Discussion}

\subsection{$Long-Term$ $Variability$}

In figure 5, the long term evolution of the two components of GS2000+25 and of GS1124-683 are shown; in figures 6 and 7, the long-term histories of various
parameters of the X-ray energy spectra are plotted. The
photon indices of the power-law component were about 2.2--3.0 until 225
d after the peak in GS 2000+25 and 121 d after the X-ray maximum (1991
January 17) in GS 1124-683, respectively. 
The hard component decreased faster than
the other component at the early phase. The second
flare (the re-flare) of the X-ray flux occurred after the hard
component has decreased to its bottom intensity (GS 1124-683).
The hard component increased again with the same photon index as that of the initial phase of the outburst.
Then, the disk component decreased and the state changed to the low state; the photon index became 1.5--1.7, which is similar to the values of other BHC-XBs in the low state, such as Cyg X-1 (Ebisawa 1991; Miyamoto et al. 1993).
Considering the value of the power spectral density and this photon index of the power-law component in the energy spectrum, the re-increase of the hard component around 121 d after the maximum of GS2000+25 is not the beginning of the low state.

The temperatures, $kT_{\rm in}$, of the innermost accretion disk dropped smoothly as the flux decreased, except for the case of the second flare up of GS 1124-683.
The time constants of the decay of $kT_{\rm in}$ of GS 2000+25 and GS 1124-683 were about 170 d and 176 d, respectively.
Radius $R_{\rm in}$ of the innermost accretion disk remained roughly constant in the early phase of the decay for both sources.
For GS 2000+25, $R_{\rm in}$ was almost constant until 225 d after the peak, and then suddenly decreased 232 d after the peak.
The temperature ($kT_{\rm bb}$) of the incident blackbody radiation photons to the Compton cloud decreased similarly to that of the inner edge of the disk ($kT_{\rm in}$), although that of GS 1124-683 was variable at the beginning of the outburst.
In the case of GS 2000+25, $kT_{\rm bb}$ took almost the same values as $kT_{\rm in}$ within a factor of 0.6--1.1, which supports the view that the seed photons of the hard component (Compton scattered component) were produced near the inner edge of the accretion disk (Miyamoto et al. 1994).
The Compton $y$-parameters of GS 2000+25 and GS 1124-683 were below 0.5 until 225 d and 121 d after the peak, respectively, and subsequently increased up to about 1.
More noteworthy is that the transition from the high state to the low state occurred between 225 and 232 d after the peak in the case of GS 2000+25, which was confirmed by changes of not only the photon index, but also the absolute values of the NPSDs of the hard component.
Thus, the transition had finished within 7 d. In the case of GS 1124-683, the transition occurred between 121 d and 148 d.

In the case of GS 2000+25, the transition from the high state to the
low state occurred about 100 days after the re-increase of the hard
component. While in the case of GS 1124-683, the transition occurred within
about 25 d after the re-increase. Thus, although the light curves of
total X-rays from these two X-ray novae are similar, the evolution of
the two energy spectral components is different from each other. In
this connection, it is of interest to note that at the beginning of
the nova GS 2000+25, the hard component was not the main component, as
long as GS 2000+25 was observed with the ASM.
In the case of GS 1124-683 the hard
component was the main component at the beginning of the nova
(Miyamoto et al. 1993; Ebisawa et al. 1994), as shown in figure 5.

Naturally, the mass-accretion rate can be considered to be the maximum
at the peak of the outburst. The peak luminosities of the two sources,
GS2000+25 and GS1124-683, were roughly 10$^{38} (\frac{D}{\rm 3kpc})^2$ erg
s$^{-1}$. Since the low-energy component can be represented by the disk
component with an inner radius of $\sim$ 10 km, it is reasonable to
understand that an optically-thick standard disk is formed. However,
a substantial hard component was observed. This hard component
was clearly different from that observed during the low state, because
the photon index was different (also short term variation is different).
When we assume the comptonized black body model for the hard
component, the source photons can be recognized as those from the
inner disk.  The high-temperature electrons may surround the inner
accretion disk like a corona. The existence of the standard disk, even
at the peak of the outburst, suggests that a transition to the
optically thick ADAF did not occur, although hot electrons must
exist around the inner accretion disk, which might be the ADAF
component co-existing with the standard disk component.
 
To explain the second and third flares during the
decline phase of the outburst, Augusteijn et al. (1993) proposed that
the intensity jumps were due to enhanced mass flow from the companion
star, which was responding, essentially linearly, to heating by X-rays
from the primary (the X-ray star), and called them echoes. They
applied their model to GS 2000+25 and could explain the X-ray light
curve with the second flare (62 d after the peak) and the third flare
(132 d after the peak) during the outburst in the 1--20 keV energy
range. According to their model, the process goes on continuously,
triggered by the initial burst, and a time delay of a few months
corresponds to the time scale of flow of the matter from the inner
Lagrangian point, L1, to the compact object. Chen et al. (1993) also
proposed the following scenario. The main outburst is caused by a disk
instability and the second maximum is due to mass transfer from the
X-ray-irradiated secondary star surface. They also explained that the
third flare observed in A0620-00 was due to a mass-transfer
instability caused by hard X-ray heating of the sub-photospheric
layer of the secondary. However, it was found that in the case of GS
2000+25 the feature of the second flare (70 d after the peak) was
different from that of the third flare 132 d after the peak, as shown
in figure 1. That is, the second and third flares were due to
increases of the soft and hard X-rays, respectively. These observational
results suggest that the causes of the second and the third flare were
different, and they can not be explained by such a simple``echoing"
mechanism.

A natural explanation of the third flare is a transition of the
disk structure from the standard disk to the optically thin ADAF
solution, which was originally pointed out by Ichimaru (1977).  The
disappearance of the disk component at the transition directly
coincides with this idea. It is also interesting to note a view by
Miyamoto et al. (1995) that X-ray flares or X-ray novae start their
X-ray flux increase in a state in which the photon index of the
power-law component of energy spectra is about 1.7 (they termed it ``the power-law-hard-state") and the source change
to the state in which the photon index of the power-law component of
about 2.7 (they termed it ``the power-law-soft-state"), near their
X-ray (1--100 keV) flux maximum, where X-ray novae have been
discovered in soft X-rays (1--20keV). This behavior also suggests that
before a soft X-ray flare the accretion disk is an optically thin
ADAF.

\subsection{$Short-Term$ $Variability$}

Assuming that the low and very high state noises (short-term variations) are one
phenomenon, van der Klis (1994a, b) concluded that in the flat-top noise
the cut-off frequency increases with the mass-accretion rate, and that the flat-top level (fractional amplitude) decreases with the mass-accretion rate. 
However, in the high state, the two energy 
spectral components (the hard component and the ultra-soft component)
should be taken into account to examine the noise. For the cases of GS2000+25 and GS1124-683, such a kind of variation of the cut-off frequency in the high state was not confirmed.
Therefore, we discuss the fitting results using a constant Gamma for the flat-top noise, as described in subsections 3.3 through 3.5.

The following points were clarified from an analysis using the NPSDs, as
mentioned in subsections 3.3, 3.4, and 3.5. (1) The variations of the observed NPSD of GS 2000+25
in the high state can be explained by the assumption that the disk
component and the power-law component have the characteristic shape of
NPSDs; that is, the power-law shape NPSD for the disk component and
the flat-top shape NPSD for the hard component, which can be
represented by a Lorenzian function of zero central frequency. Moreover, the normalized 
flat-top level of the power-law component is constant during the high state, as shown in figures 10 and 11. This is consistent with the conclusion obtained in GS 1124-683 by Miyamoto
et al. (1994).  (2) In the high state, the NPSDs of the hard component
are larger than those of the disk component by a factor of about 70 at
0.3 Hz. The NPSDs in the low state are still larger than those of the
hard component in the high state by a factor of more than 10, which is
also consistent with the conclusion which Miyamoto et al. (1994)
obtained for GS 1124-683.  (3) The values of the characteristic NPSD of
the two components of GS 2000+25 are similar to those of GS 1124-683.

Also, it should be noted that the hard components of GS 2000+25 and GS
1124-683 in the low state are quite different from those in the
high state with respect to not only the photon index of the power-law
energy component (or the Compton $y$-parameter of the Compton scattering
component), but also to the values of the NPSDs. 

If we assume that the accretion disk at the low state is an optically
thin ADAF, although it is a stable solution, it must have a highly
variable structure.  From this point of short-term variability in the low state, the other branch, a gas-pressure
supported and radiative cooled disk may also be possible, since it is
thermally unstable and spontaneously produces the complex structure of
the accretion disk.

\subsection{$Bimodal$ $Behavior$ $of$ $the$ $BHC-XBs$}

Remarkable distinctions between the high and low states were
found not only in the energy spectra, but also in the Normalized Power
Spectral Density (NPSD) of GS 2000+25 and GS 1124-683. Thus, the relations
between the photon index of the power-law component and the value of
NPSD were investigated for other BHC-XBs: Cyg X-1, LMC X-3, GX 339-4,
GS 1826-24, GS 1354-64 and GS 2023+338, in addition to GS 2000+25 and
GS 1124-683. For this investigation, we used data observed with the
Ginga satellite. The observation logs are given in table 8. The
relations between the photon indices and the values of the NPSDs at 0.3 Hz
are shown in figure 12. Because GS 2023+338 showed rapid variabilities in
its X-ray energy spectra and in the NPSDs due to the partial
absorption by cold matter (Oosterbrek et al. 1997), only the data which did not show partial absorption were used.

It is recognized that all of the data can be clearly divided between the
high and low states; In the low state, the photon index is
1.4--1.8 and the values of NPSDs at 0.3 Hz are about 10$^{-2}$. In the
high state, the photon index is 2.1--3.0 and the values of the NPSDs at
0.3 Hz are 10$^{-5}$--10$^{-3}$; i.e. both the energy spectra and the NPSDs are
different between the high and low states. More noteworthy is
that there is a ``gap" between the high and low states on the NPSD
vs the photon index diagram in spite of various occasions to observe eight sources.
The separation between the very high state and the high quiet state, 
which are sub-states of the so-called high state, could be approximately 
given by values of approximately 10$^{-4}$ for the NPSD at 0.3 Hz.

The correlation between the PLF and the value of NPSD at 0.3 Hz for
various sources in various states is shown in figure 13. Although some
exceptions exist, the same correlation as mentioned in subsection 3.4 has been
confirmed for the other BHC-XBs. The fitting result for all data in
the high state with equation (3) in subsection 3.4 is also shown in this
figure by the thick line. The most probable values of NPSD$_{\rm hard}$(0.3) and
NPSD$_{\rm disk}$(0.3) in the high state are (7.7$\pm$0.1)$\times$10$^{-4}$ and (1.7$\pm$0.6) $\times$10$^{-5}$
(reduced $\chi ^2$=15). Thus, it is clear that the solid line based on the
fitting result approximately represents the relation between the PLF
and the value of NPSD at 0.3 Hz in the high state (both the very high state
 and the high quiet state) within a factor of
about 4, although its reduced $\chi ^2$ is large. Miyamoto et al. (1992)
pointed out that the NPSDs of various sources in the low state are
almost the same, which correspond to that the data of which the PLF
are above 0.9. This is also confirmed in figure 13, and the mean value
of NPSD$_{\rm hard}$(0.3) in the low state is about 2.5$\times$10$^{-2}$. Thus,
similarities of the NPSDs of the various BHC-XBs are shown not only in
the low state, but also in the high state.

Nowak (1994) assumed that the NPSD$_{\rm power law}$ and the NPSD$_{\rm flat
top}$ were due to viscous instabilities and the thermal instability of
the accretion disk, respectively, and calculated NPSD in the
very high state (a sub-state of the high state, where the power-law
component is dominant). He showed in figure 1 of his paper (Nowak
1994) that the flat-top level of the NPSD increased with the mass-accretion rate. This could be consistent with our observations of
GS2000+25 and GS1124-683 shown in figure 10. However, he did not
calculate the amounts of the hard and ultra-soft components. We thus
could not estimate the values of PLF and compare his results directly
with the values shown in figure 10 and 11. He also showed a large
change of the cut-off frequency of the NPSD$_{\rm flat top}$, which seemed
not to be consistent with our observation. Nowak (1994) also mentioned
that the very high state is a transition phase between the high and low states. This can be true because Miyamoto et al. (1995)
showed that the black hole X-ray nova in its rising phase increases
its X-ray intensity in the power-law hard state (the low state), and at
its near maximum it changes to the power-law soft state (the high
state).

\section{Summary}

The X-ray energy spectra and NPSDs of GS 2000+25 were investigated. These
results indicate that there were, at least, three X-ray emission
mechanism during the 1988 outburst of GS 2000+25: the ultra-soft component
in the high state, the hard component in the high state, and the hard
component in the low state. The ultra-soft component in the
high state could be represented by radiation from an optically-thick
accretion disk (the disk component).  The hard component in the
high state had a power-law shape, which could be explained by Compton-scattered radiation, of which the Compton $y$-parameter was 0.3-0.5. 
The incident photons to the Compton cloud are emitted from the central
part of the accretion disk. The values of the NPSD of the hard
component in the high state were larger than those of the disk
component by factor of about 50--70 at 0.3 Hz. The hard component
in the low state had a power-law shape of which the photon index ranged from 1.4 to 1.8.
The values of the NPSD of the hard component in the low state were
still larger than those of the hard component in the high state by a
factor of more than 10. Thus, the hard components in the high and low states in essence differed from each other.
 
We can speculate our results according to the current accretion-disk
models as follows: (1) Even in the peak of the outburst, the standard
disk exits. (2) There are a high-temperature electrons around the inner
disk, which may relate to the optically thick ADAF. (3) The optically
thin ADAF in the low state is very variable, and not a steady flow.

These features of three components are common to other BHC-XBs, such
as Cyg X-1, GX 339-4, GS 1124-683, LMC X-3.

\acknowledgments

We wish to thank Prof. Miyamoto.  Many of the ideas in this paper originated from him. Large parts of this paper were completed with his help and his encouragement. We also thank a group who supported this
work throughout the observations and data analysis. We also
acknowledge helpful discussions with Prof. H. Tsunemi. We thank 
Dr. K. Hayashida for his assistance in constructing the background
spectrum to derive the energy spectrum of GS 2000+25. Takizawa's (1991)
analysis of the X-ray energy spectra of GS 2000+25 was very
valuable for investigating short term variability of X-rays in the
high state and the low state. We thank also to Dr. Michael Morris and
an anonymous reviewer for reading the draft and making a number of
valuable suggestions . This study was partly supported by the
Scientific Research Grant-in-Aid of the Ministry of Education, Culture,
Sports Science and Technology (Nos. 07740189, 07222210 and 08740173).

\section*{References}

Abramowicz, M. A., Chen, X., kato, S., Lasota, J.-P., \& Regev, O. 1995, ApJ, 438, L37

Abramowicz, M. A., Czerny, B., Lasota, J.-P., \& Szuszkiewicz, E. 1988, ApJ, 332,646

Augusteijn, T., Kuulkers, E., \& Shaham, J. 1993, A\&A, 279, L13

Baym, G., \& Pethick, C. 1979, ARA\&A, 17, 415

Barret, D., McClintock, J. E., \& Grindlay, J. E. 1996, ApJ, 473, 963

Beekman, G., Shahbaz, T., Naylor, T., \& Charles, P. A. 1996, MNRAS, 281, 1

Belloni, T., \& Hasinger, G. 1990, A\&A, 227, L33

Belloni, T., M\'endez, M., van der Klis, M., Hasinger, G., Lewin, W. H. G., \& van Paradijs, J. 1996, ApJ, 472, L107

Belloni, T., M\'endez, M., van der Klis, M., Lewin, W. H. G., \& Dieters, S. 1999, ApJ, 519, L159

Belloni, T. 2001, astro-ph/0112217

Callanan, P. J., \& Charles, P. A. 1991, MNRAS, 249, 573

Callanan, P. J., Garcia, M. R., Filippenko, A. V., McLean, I., \& Teplitz, H. 1996, ApJ, 470, L57

Casares, J., Charles, P. A., \& Marsh, T. R. 1995, MNRAS, 277, 45

Charles, P. A., Kidger, M. R., Pavlenko, E. P., Prokofieva, V. V., \& Callanan, P. J. 1991, MNRAS, 249, 567

Chen, W., Livio, M., \& Gehrels, N. 1993, ApJ, 408, L5

Cominsky, L., Jones, C., Forman, W., \& Tananbaum, H., 1978, ApJ, 224, 46

Ebisawa, K. 1991, PhD Thesis, The University of Tokyo

Ebisawa, K., Ogawa, M., Aoki, T., Dotani, T., Takizawa, M., Tanaka, Y., Yoshida, K., Miyamoto, S., et al. 1994, PASJ, 46, 375

Esin, A. A., Narayan,R., Cui, W., Grove, J. E., \& Zhang, S. -N. 1998, ApJ, 505, 854

Filippenko, A. V., Matheson, T., \& Barth, A. J. 1995, ApJ, 455, L139

Frank, J., King, A., \& Raine, D. 1992, Science, 258, 1015

Garcia, M. R., McClintock, J. E., Narayan, R., Callanan, P., Barret, D., \& Murray, S. S. 2001, ApJ, 553, L47

Gilfanov, M., Churazov, E., \& Sunyaev, R. 1999, in Observational Evidence for Black Holes in the Universe, ed. S. K. Chakrabarti (Dordrecht: Kluwer), 319

Ichimaru, S. 1977, ApJ, 214, 840

Kato, S., Fukue, J. \& Mineshige, S. 1998, in Black-Hole Accretion Disks, ed. S. Kato, J. Fukue, \& S. Mineshige. (Kyoto: Kyoto University Press)

Kitamoto, S. 1999, in Observational Evidence for Black Holes in the Universe, ed. S. K. Chakrabarti (Dordrecht: Kluwer), 369

Kitamoto, S., Mizobuchi, S., Yamashita, K., \& Nakamura, H. 1992a, ApJ, 384, 263

Kitamoto, S., Tsunemi, H., \& Roussel-Dupre D. 1992b, ApJ, 391, 220

Liang, E. P., \& Nolan, P. L. 1984, Space Sci. Rev. 38, 353

Makino, F., 1987, Astrophys. Lett. Comm. 25, 223

Makishima, K. 1988, in Physics of Neutron Stars and Black Holes,
ed. Y. Tanaka (Tokyo: Universal Academy Press), 175

McClintock, J. E., \& Remillard, R. A. 1986, ApJ, 308, 110

Mineshige, S. Hirose, M., \& Osaki, Y. 1992a, PASJ, 44, L15

Mineshige, S. Ebisawa, K., Takizawa, M., Tanaka, Y., Hayashida, K., Kitamoto, S., Miyamoto, S., \& Terada, K. 1992b, PASJ, 44, 117

Mitsuda, K. Inoue, H., Koyama, K., Makishima, K., Matsuoka, M., Ogasawara, Y., Shibazaki, N., Suzuki, K., Tanaka, Y., \& Hirano, T. 1984, PASJ, 36, 741

Miyamoto, S., Kitamoto, S., Mitsuda, K., \& Dotani, T. 1988, Nature, 336, 450

Miyamoto, S., Kimura, K., Kitamoto, S., Dotani, T., \& Ebisawa, K. 1991, ApJ, 383, 784

Miyamoto, S., Kitamoto, S., Iga, S., Negoro H., \& Terada, K. 1992, ApJ, 391, L21

Miyamoto, S., Iga, S., Kitamoto, S., \& Kamado, Y. 1993, ApJ, 403, L39

Miyamoto, S., Kitamoto, S., Iga S., Hayashida, K., \& Terada, K. 1994, ApJ, 435, 398

Miyamoto, S., Kitamoto, S., Hayashida, K., \& Egoshi, W. 1995, ApJ, 442, L13

Nishimura, J., Mitsuda, K., Itoh, M. 1986, PASJ, 38, 819

Nolan, P. L. Gruber, D. E., Matteson, J. L., Peterson, L. E., Rothschild, R. E., Doty, J. P., Levine, A. M., Lewin, W. H. G., \& Primini, F. A. 1981, ApJ, 246, 494

Nowak, M. A. 1994, ApJ, 422, 688

Oda M. 1977, Space Sci. Rev., 20, 757

Oosterbroek, T., van der Klis, M., van Paradijs, J., Vaughan, B., Rutledge R., Lewin, W. H. G., Tanaka, Y., Nagase, F. et al. 1997, A\&A, 321, 776

Shakura, N. I., \& Sunyaev, R. A. 1973, A\&A, 24, 337

Takizawa, M. 1991, Master Thesis, The University of Tokyo (in Japanese)

Tanaka, Y. 1989, in Proc. 23rd ESLAB Symposium (ESA SP-296) Vol.1, ed J. Hunt, B. Battrick (Noordwijk, ESA), 3

Tanaka, Y., \& Shibazaki, N. 1996, ARA\&A, 34, 607

Terrell, N. J., Jr. 1972, ApJ, 174, L35

Tsunemi, H., Kitamoto, S., Okamura, S., \& Roussel-Dupre, D. 1989a, ApJ, 337, L81

Tsunemi, H., Kitamoto, S., Manabe, M., Miyamoto, S., Yamashita, K., \& Nakagawa, M. 1989b, PASJ, 41, 391

Turner, M. J. L. Thomas, H. D., Patchett, B. E., Reading, D. H., Makishima, K., Ohashi, T., Dotani, T., Hayashida, K. et al. 1989, PASJ, 41, 345

van der Klis, M., Hasinger, G., Stella, L., Langmeier, A., van Paradijis, J., \& Lewin, W. H. G. 1987, ApJ, 319, L13

van der Klis, M. 1994a, A\&A, 283, 469

van der Klis, M. 1994b, ApJS, 92, 511

Watarai , K., Mizuno, T., \& Mineshige, S. 2001, ApJ, 549, L77

Weisskopf, M. C., Kahn, S. M., \& Sutherland, P. G. 1975, ApJ 199, L147

White, N. E., Kaluzienski, J. L., \& Swank, J. H. 1984, in AIP conf. Proc. 115, High Energy
Transients in Astrophysics, ed. S. E. Woosley (New York: AIP), 31

White, N. E., \& Marshall, F. E. 1984, ApJ, 281, 354

\clearpage

\section*{Figure Legends}
\noindent
Fig. 1. X-ray light curves of GS 2000+25 and its hardness ratio
observed with the ASM onboard the Ginga satellite. In the upper panel,
the data in the 1--6 keV and 6--20 keV energy ranges are shown; their
ratios are shown in the lower panel.\\

\noindent
Fig. 2. X-ray Light curve (1.7--37 keV) of GS 2000+25 observed with
the LAC.\\

\noindent
Fig. 3. X-ray energy spectra of GS 2000+25 from 1988 April 30 to
December 16.\\

\noindent
Fig. 4. Fitting results of energy spectra and its residual from 1988
April 30 to December 16. The adopted model is model 2 (the
disk component and the power-law component with a smeared edge).\\

\noindent
Fig. 5. Time evolution of X-ray flux of GS 2000+25 (a) and GS 1124-683
(b) in the 1.7--37 keV energy range.  The results of model 2 were used.\\

\noindent
Fig. 6. Time evolution of the best-fit values of various
parameters of the energy spectral fitting.  ($\bullet$, GS
2000+25; $\circ$, GS 1124-683).  Here, we used model 2 (the disk
component and the power-law component with a smeared edge).\\

\noindent
Fig. 7. Time evolution of the best-fit values of various
parameter of the energy spectral fitting ($\bullet$, GS 2000+25; $\circ$, GS 1124-683), where the adopted model is model 3 (the disk component and the Comptonized blackbody component).\\

\noindent
Fig. 8. Comparisons of the NPSDs of GS 2000+25 with those
of GS 1124-683 and Cyg X-1. The thin solid line is the NPSD of GS
1124-683 obtained on 1991 January 11 (PLF=0.80). The dotted line is
the NPSD of GS 1124-683 obtained on 1991 February 26 (PLF=0.066). The
broken line is the NPSD obtained on 1991 June 13 and July 22 when GS
1124-683 was in the low state (PLF=0.95 and 0.99). The thick solid line
is the NPSD of Cyg X-1 in the low-state (on 1990 May 10). These NPSDs
were the fitting results of raw NPSD data.\\

\noindent 
Fig. 9. Fitting results of NPSDs of GS 2000+25 by two functions: the power-law-shape function and the flat-top-shape
function (Lorenzian function). The dotted line, the broken line and
the thin line are PSD$_{\rm power law}$, PSD$_{\rm flat top}$, and their sum. The
thick line in the figure on December 16 is the NPSD of Cyg X-1 in the
low state on 1990 May 9.\\

\noindent
Fig. 10. Relations between the PLFs and the NPSDs at 0.3
Hz of GS 2000+25 (filled circle) and GS 1124-683 (open square). The
most probable fitting results with equation (3) in subsection 3.4 except
for the data in the low state, are shown by the solid line for GS
2000+25 and the broken line for GS 1124-683 (Miyamoto et
al. 1994). The values of the NPSD$_{\rm hard}$(0.3) and the NPSD$_{\rm disk}$(0.3) of GS
2000+25 are (1.59$\pm$0.06)$\times$10$^{-3}$ and (2.4$\pm$1.0)$\times$10$^{-5}$, respectively (reduced
$\chi ^2$=3.7). The values of the NPSD$_{\rm hard}$(0.3) and the NPSD$_{\rm disk}$(0.3) of GS
1124-683 are (9.4$\pm$0.9)$\times$10$^{-4}$ and (1.36$\pm$0.53)$\times$10$^{-5}$, respectively
(Miyamoto et al. 1994).\\

\noindent 
Fig. 11. Relations between the Normalized PSD$_{\rm power law}$ (the
NPSD$_{\rm disk}$) and the Normalized PSD$_{\rm flat top}$ (the NPSD$_{\rm hard}$) and the
PLFs. The symbols of filled circles and open squares mean the data of the GS
2000+25 and of the GS 1124-683, respectively. In the high state (PLF $<$ 0.61),
the mean values of the PSD$_{\rm power law}$ and PSD$_{\rm flat top}$ at 0.3 Hz
normalized by the disk component and the power-law component of GS
2000+25 are (2.7$\pm$0.6)$\times$10$^{-5}$ and (9.7$\pm$1.8)$\times$10$^{-4}$, respectively, assuming
the values are constant (broken lines).\\

\noindent 
Fig. 12. Relations between the photon-index of the
power-law component and the values of the NPSDs at 0.3 Hz for various
BHC-XBs.\\

\noindent
Fig. 13. Relations between the PLFs and the values of the NPSDs at
0.3 Hz of various BHC-XBs. The fitting result for all of the data in
the high state with equation (3) in subsection 3.4 is also shown in this
figure by the solid line. The most probable values of NPSD$_{\rm hard}$(0.3) and
NPSD$_{\rm disk}$(0.3) in the high-state are (7.7$\pm$0.1)$\times$10$^{-4}$ and (1.7$\pm$0.6)$\times$10$^{-5}$
(reduced $\chi ^2$=15).\\

\end{document}